\documentclass[12pt]{iopart}
\usepackage{epsf,iopams,graphicx}  
\def\msun{$M_{\odot}$}

\def\MNRAS{{\em Mon. Not. Roy. Astron. Soc. }}
\def\lsim{\mathrel{\rlap{\lower4pt\hbox{\hskip1pt$\sim$}}
    \raise1pt\hbox{$<$}}}

\begin{document}

\title[A time-frequency method to detect EMRIs]{Detecting extreme mass ratio inspirals with LISA using time-frequency methods}

\author{Linqing Wen\dag \footnote[3]{email address: lwen@aei.mpg.de},
 Jonathan R Gair\ddag}
\address{\dag Max Planck Institut fuer Gravitationsphysik, Albert-Einstein-Institut
Am Muehlenberg 1,  D-14476 Golm, Germany}
\address{\ddag  Theoretical Astrophysics, California Institute of Technology, Pasadena, CA 91125 and Institute of Astronomy, University of Cambridge, Madingley Road, Cambridge, CB3 0HA, UK}
\begin{center}
\begin{abstract}
The inspirals of stellar-mass compact objects into supermassive black
holes are some of the most important sources for LISA.  Detection
techniques based on fully coherent matched filtering have been shown
to be computationally intractable. We describe an efficient and robust
detection method that utilizes the time-frequency evolution of such
systems. We show that a typical extreme mass ratio inspiral (EMRI)
source could possibly be detected at distances of up to $\sim 2$ Gpc,
which would mean $\sim$ tens of EMRI sources can be detected per year using this technique.  We discuss the feasibility of using this method as a first step in a hierarchical search.
\end{abstract}
\end{center}



\maketitle

\section{Introduction}
\label{intro}
Astronomical observations indicate that many galaxies host a supermassive black hole (SMBH) in their centre. The inspirals of stellar-mass compact objects into such SMBHs with mass $M\sim\mbox{few}\times10^{5}M_{\odot}$--$10^{7}M_{\odot}$ constitute one of the most important gravitational wave (GW) sources for the planned space-based GW observatory LISA.  Preliminary results \cite{jon04}  indicate that the LISA EMRI detection rate will most likely be dominated by inspirals of $\sim10$ \msun\ BHs into $\sim10^6$\msun\ SMBHs. The EMRI detection rate could be as many as $\sim 1000$ in 3--5 years within $\sim 3.5$ Gpc.  

The strain amplitude of GWs from EMRIs can be estimated using the Newtonian quadrupole approximation to the Einstein field equations, 
\begin{equation}
h \sim 6 \times 10^{-22} \left (\frac{d}{\mbox{Gpc}}\right )^{-1} \left (\frac{M}{10^6 M_\odot}\right )^{2/3}
\frac{\mu}{10 M_\odot} \left (\frac{f}{5 \mbox{mHz}}\right )^{2/3},
\end{equation}
where $f$ is the orbital frequency, $d$ is the distance of the source 
from the Earth and $\mu =mM/(m+M)$ is the reduced mass. This can be 
compared with the characteristic noise strain of $\sim 5 \times 10^{-21}$ 
at the floor of the LISA noise curve near 5 mHz \cite{curt98, leor04}.  For a $10 + 10^6 $\msun\  EMRI system at 1 Gpc, the instantaneous signal-to-noise ratio (SNR) $\rho_t$ is at best around 0.1.  Detection of GWs from EMRIs therefore depends on (semi-) coherent accumulation of the signal with time.   

The optimal method to detect a known time series signal $h(t)$ embedded in stationary Gaussian noise $n(t)$ is matched filtering. In that technique,  we search for the maximum correlation of the Fourier components of the data with that of the known waveforms, weighted by the noise variance. The optimal SNR, $\rho_M$, can be written as
\begin{equation}
\rho^2_M =\sum^N_{k=1} \frac{2h^2_k}{\sigma^2_{n_{k}}},
\end{equation}
\label{MF}
where $h_k$ is the Fourier amplitude of the signal,
$\sigma^2_{n_k}=0.5S_{h}(f)/ (\rmd t^2\rmd f) $ is the expected
variance of the noise component $n_k$ at frequency bin $k$,
characterized by $S_h(f)$, the
strain spectral density of the noise, $N$
is  the number of Fourier frequency bins and $\rmd f$ is the bin
width. The SNR squared is therefore effectively proportional to the
product of the number of wave cycles with the instantaneous SNR
squared. During an integration over the lifetime of LISA
($\sim3$--$5$ yrs), the number of GW cycles observed, $N_{GW}\sim Tf
\sim 5\times 10^5$, so the optimal SNR can be as high as
$\rho_M \sim 100$ at 1 Gpc. 

\section{Computational challenges of EMRI detection}
EMRI waveforms are complex and are characterized by many frequency
components, which arise from several effects. First, typical EMRI
orbits are expected to be still moderately eccentric, $e\sim0$--$0.5$,
during the last several years of inspiral when LISA can detect them
\cite{leor04}--\cite{sigurdsson97}. At such moderate eccentricities,
there can be as many as five harmonics of the orbital frequency contributing significantly ($>10 \%
$) to the observed SNR \cite{peters63}. In addition, EMRI signals exhibit many modulations, caused by periastron precession, spin induced precession of the
orbital plane and yearly amplitude and Doppler modulation due to the motion of LISA around the sun. Finally, the frequency components in an EMRI signal exhibit significant evolution over a LISA observation. For a 3 year observation of a signal with central frequency $\sim5$ mHz, the signal power can be spread over as many as $10^5 $ frequency bins \cite{leor04}. This hinders the detection of the signals using simple Fourier spectrum analysis.

The complexity of the EMRI waveforms makes a fully coherent matched
filtering search computationally impossible. Rough estimates would
suggest that $\sim10^{40}$ templates are needed for a fully coherent
search \cite{jon04}. Extrapolating to the time of the LISA mission, it
is reasonable to assume $\sim50$ Tflops of available computing power for the search, but this allows only $\sim10^{12}$ templates to be searched in real time. Alternative methods are therefore required to detect EMRIs, such as semi-coherent hierarchical searches \cite{jon04}.

\section{A time-frequency detection method}
We describe an efficient and robust strategy to detect GWs from EMRIs by accumulating the signal power in the time-frequency (t-f) domain. The t-f power spectrum is produced by dividing the data into 2 week long segments and carrying out a Fast Fourier Transform (FFT) on each. In the semi-coherent matched filtering search \cite{jon04}, the waveform is also divided into sections, of $\sim3$ weeks. In that case, this is the longest segment length that computational constraints will allow. In the time-frequency analysis, there are no such computational limits, but we choose a 2-week duration to ensure enough time and frequency resolution to trace the frequency evolution of EMRIs with time. The power spectrum is defined for each segment $i$ and frequency bin $k$ as,
\begin{equation}
P(i,k)= \frac{2|(h^{i}_k+n^{i}_k)|^2}{\sigma^2_{n_k}}=\frac{2
(h^{i}_k)^{2}}{\sigma^2_{n_k}}+4\frac{\mbox{Re}[h^{i}_k (n^{i}_k)^{*}]} {\sigma^2_{n_k}} +\frac{2 (n^{i}_k)^{2}}{\sigma^2_{n_k}}.
\label{power}
\end{equation}
We then calculate the power ``density'', $\rho(i,k)$, by computing the average power within a rectangular box centered at each point ($i,k$), 
\begin{equation}
\rho(i,k)= \sum^{n/2}_{a=-n/2} \sum^{l/2}_{b=-l/2} P(i+a,k+b)/m,
\end{equation}
where $n$, $l$ are the lengths of the box in the time and frequency
dimension respectively and $m=n\times l$ is the number of data points in the box.   The SNR at each point $(i,j)$ is then $\rho_s = (\rho-\bar{\rho})/\sigma_\rho$, where $\bar{\rho}$ is the mean of $\rho$ calculated in the entire t-f plane and $\sigma^2_\rho$ is the expected variance of $\rho$ for pure noise. In practice, we use the variance of the calculated $\rho$ in the entire t-f plane.  The detection process involves finding the local maximum $\rho_s$ or tracks of ``excess'' $\rho_s$.  

If the data consist of only stationary Gaussian noise, $m\rho$ will
follow a $\chi^2_{2m}$ distribution, with expected $\sigma_\rho ={2}/{\sqrt{m}}$, i.e., the larger the box, the smoother the noise power density in the t-f plane. For a given box size, the false alarm
probability (FAP) for finding at least one point with $\rho_s$ above a certain threshold $\rho_0$ is 
\begin{equation}
FAP_m \sim N_f Q_{\chi^2_{2m}} (\sqrt{4m}\rho_0+2m), 
\label{fapmEq}
\end{equation}
where $Q_{\chi^2_{2m}} (P)$ is the cumulative distribution function for the
$\chi^2_{2m}$ distribution. We estimate $N_f \sim N/(m/4)$ for the number of
independent data points searched.

To search for a possible signal, we vary the box lengths $n$ and $l$ until the maximum (or a significant) $\rho_s$ is found. 
The optimal box size should be large enough to contain most of the signal power but small enough to exclude most of the noise contribution.  The overall probability of finding a FAP$_m$ below some threshold FAP$_0$ depends on the number of independent trials of different box sizes. A Monte Carlo simulation is in progress to determine the statistics of this method and to compute appropriate thresholds. In the present work, the FAP of the search is based on a simple case where we increase the box dimensions by factors of two, one side at a time, and the overall FAP is estimated as FAP$_m$ multiplied by the number of boxsizes searched. In this paper, significant detections are defined as those such that the overall FAP of the search is $<10^{-2}$.

Like many other time-frequency signal processing methods, this method examines the statistics of the presence of a lot of high power in a region. Our method is in particular similar to the ``excess power'' method \cite{anderson01}, as both use the summation of powers within a certain time and frequency interval.  The excess power method was designed to detect bursting waveforms. Our approach applies to the detection of both burst-like and continuous waves since it helps to map out the structure of the excess power density. This structure can then be detected by finding the local maximum or using pattern-recognition methods.

\section{Simulated EMRI waveform}
To test this approach, we tried to detect an EMRI signal in simulated
data. Accurate inspiral waveforms are not yet available, so we made
use of approximate numerical waveforms, as described in
\cite{jon04,kostas02,tev04}. We considered a ``typical'' EMRI event
--- the inspiral of a $10$\msun\ BH into a $10^6$\msun\ SMBH, with
eccentricity $e=0.4$ and pericentre $r_{p} \approx 11M$ at the start
of the observation, SMBH spin of $a=0.8M$, orbital inclination angle
of $45^{{\rm o}}$ (using the definition of inclination in
\cite{kostas02}) and placed at distances of $0.5$--$2$ Gpc. We used
data of total duration three years, sampled at a cadence of $8$s. With these choices, the total number of data points analyzed was $N=1.2\times 10^7$.  The simulated data consist of two independent LISA
data streams (the low frequency `I' and `II' responses described in
\cite{curt98}). The combined matched filtering SNR at a distance of
$1$ Gpc is $\rho_M \sim 140$ for the whole three years of data, and $\sim 90$ for the last year. We used the LISA noise response given in \cite{leor04}.


\section{Results and Discussion}
\label{reslt}
\begin{figure}
\centerline{\includegraphics[keepaspectratio=true,height=6.5in,angle=0]{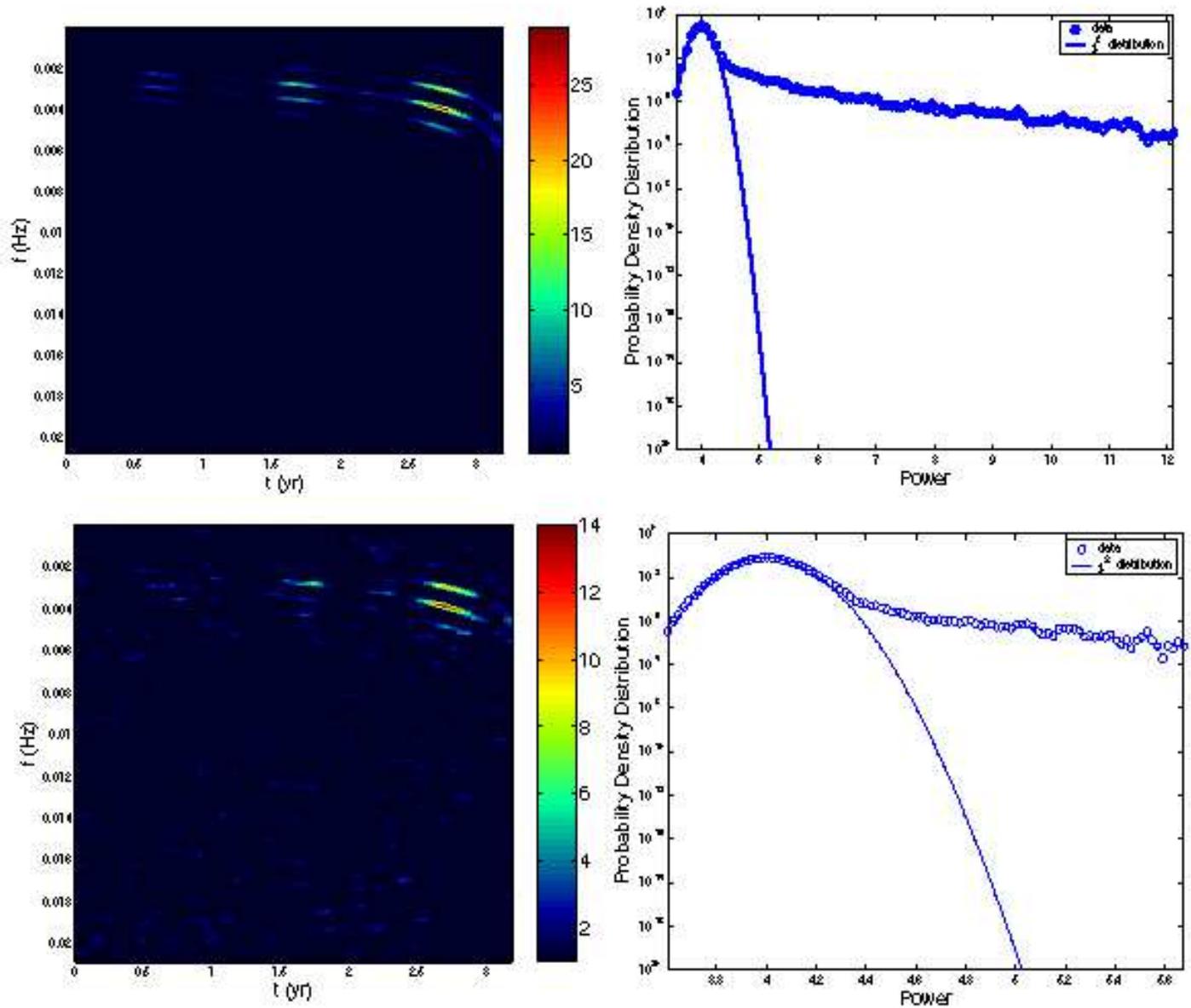}}
\caption{Left -- the t-f (normalised) power density for the optimal box 
size. Right -- the distribution of power (circles) plus expected 
distribution for pure noise (solid line). The upper plots are for $d=0.5$ 
Gpc (optimistically, we expect $\lsim3$ such events in three years). This 
could be detected at a FAP of $<10^{-16}$ and a maximal SNR of $\sim28$. 
The lower plots are for $d=1$ Gpc (we expect $\lsim25$ events in three 
years) and have FAP$<10^{-16}$ and maximal SNR$\sim14$.}
\label{fig1}
\end{figure}

\begin{figure}
\centerline{\includegraphics[keepaspectratio=true,height=6.5in,angle=0]{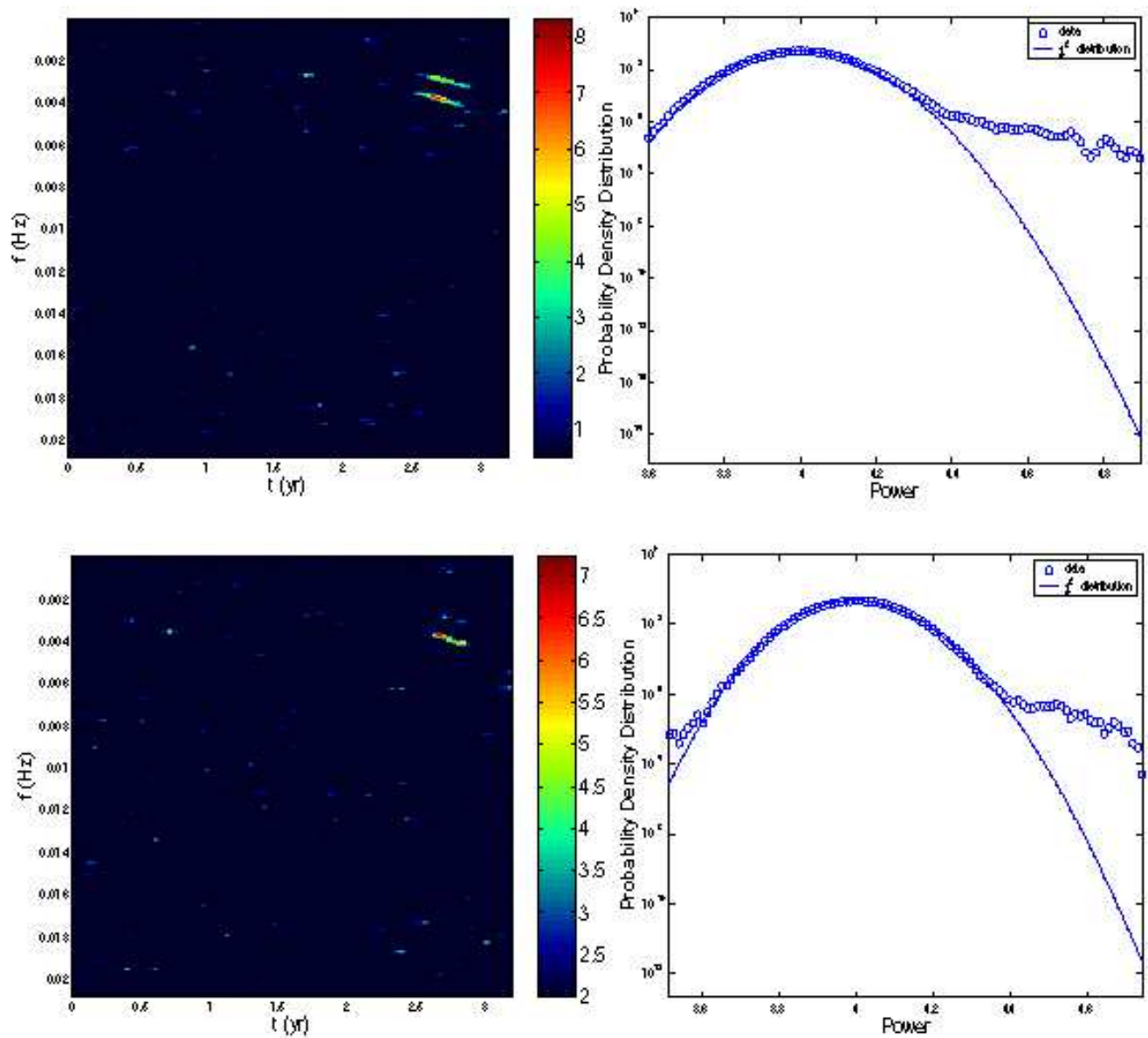}}
\caption{As Figure~\ref{fig1} but for $d=1.4$ Gpc (upper plots, expect $\lsim60$ events in three years, FAP$<10^{-10}$ and SNR$_{max}\sim8$) and $d=2$ Gpc (lower plots, expect $\lsim180$ events in three years,  FAP$\sim2\times10^{-6}$ and SNR$_{max}\sim7$).}
\label{fig2}
\end{figure}

In Figures~\ref{fig1} and \ref{fig2} we show the normalized power density $\rho_s$ in the time-frequency domain calculated with the ``optimal'' box size when the EMRI was at a distance of $0.5$, $1.0$, $1.4$, and $2$ Gpc respectively. We also show the power distribution function and the pure noise theoretical expectation for comparison.  

At the distance of $0.5$ and $1$ Gpc, the evolution of the GW central
frequency (and harmonics) with time is apparent to the eye in the
time-frequency plane. The amplitude increases as the particle
inspirals but the signal is also  modulated by LISA's motion. At $0.5$
Gpc, GWs from the last year of inspiral can be detected at SNR $\sim
28$, $19$, and $8$, respectively at each of the three dominating frequency
components.  At the distance of $1.4$ Gpc, the frequency evolution is
visible over the last year and two frequency components are apparent. At a distance of $2$ Gpc, the signal can possibly be detected with an SNR of $\sim 7$, and an overall FAP of $\sim 2\times 10^{-6}$ when searching through all independent trials.

\begin{figure}
\centerline{\includegraphics[keepaspectratio=true,height=6.5in,angle=90]{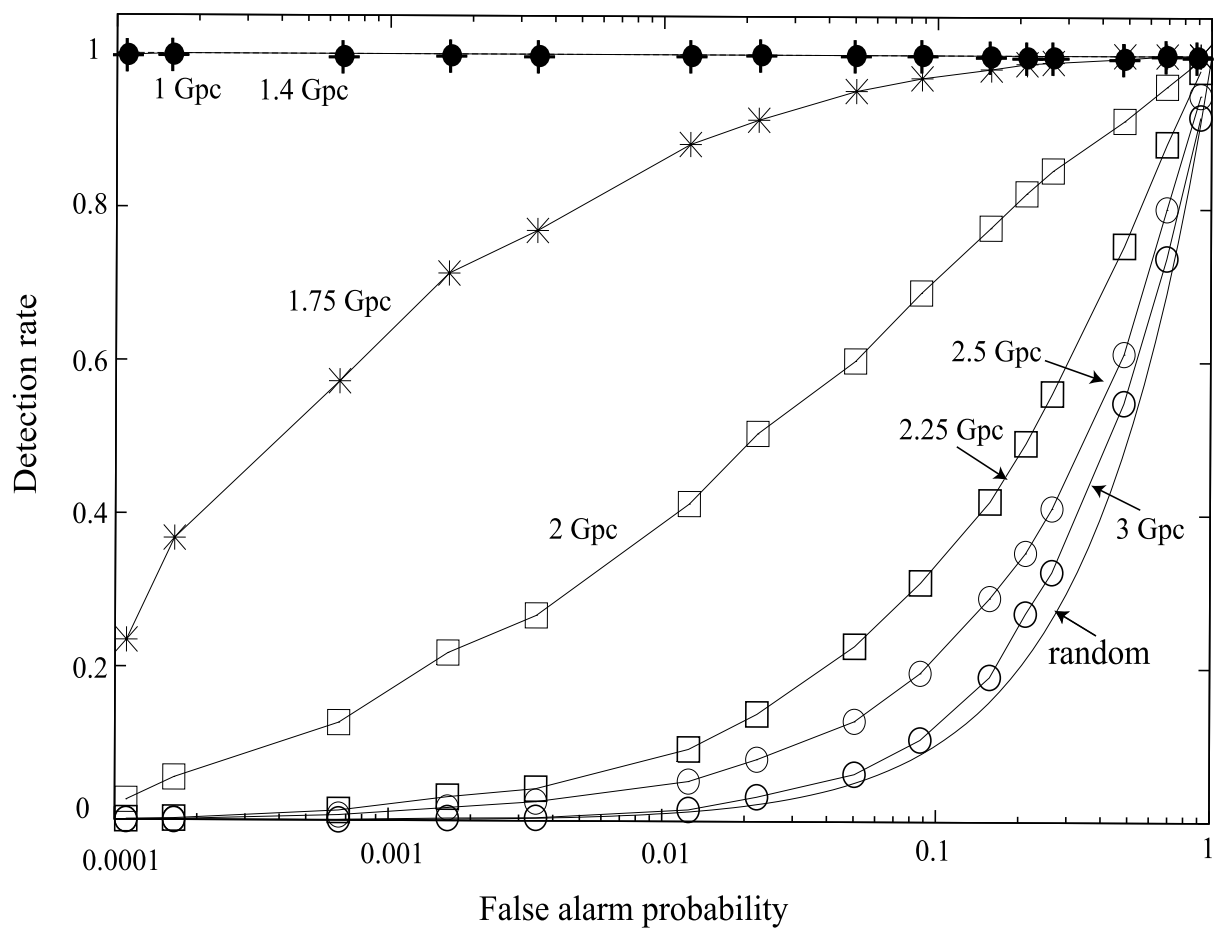}}
\caption{Approximate ROC curve for this method. The detection rate is shown as a function of the overall false alarm probability of the search, when the source is placed at distances of $1$, $1.4$, $1.75$, $2$, $2.25$, $2.5$ and $3$ Gpc from the detector. The performance of a random search, for which the false alarm rate equals the detection rate, is shown for comparison.}
\label{fig3}
\end{figure}

To assess the efficiency of this method, we show in Figure~\ref{fig3} an approximate Receiver Operator Characteristic (ROC) curve for this method. The ROC is
shown for the sources at $1$ Gpc, $1.4$ Gpc and $2$ Gpc discussed in the text, and
also distances of $1.75$ Gpc, $2.25$ Gpc, $2.5$ Gpc and $3$ Gpc for comparison. The
ROC curves were computed by setting thresholds on $\rho$ for each bin
size and performing a preliminary Monte Carlo of $\sim20000$ noise realisations. The false
alarm probability was computed as the fraction of pure noise realisations
in which a threshold was exceeded for at least one bin size. The detection
rate was the fraction of realisations of signal plus noise in which the
maximum SNR exceeded the threshold for at least one bin size. The thresholds were set by fixing the $FAP_m$ defined by equation (\ref{fapmEq}) to be equal for all bin sizes, taking $N_f = N/(m/4)$. Different choices of thresholds amount to distributing the overall FAP of
the search between the various bins in different ways. The optimum
threshold choice for a single source will be source dependent. Monte Carlo
simulations are underway in order to optimise the threshold choice in the
sense of giving the best performance. We see that the detection
performance is very good up to $1.75$ Gpc. At $2$ Gpc, the detection rate is
still in excess of $50\%$ for an overall false alarm probability of a few
percent. The source at $3$ Gpc represents the absolute limit of this
particular search, since that is the point at which this search ceases to
do any better than a random one. This should be contrasted with the
performance of the semi-coherent matched filtering technique
\cite{jon04}. An ROC curve is not available for that algorithm, but based
on the results of Gair et al., at an overall false alarm probability of
$1\%$ the detection rate for this source at a distance of $2$ Gpc would likely be close to $100\%$. However, as emphasised before, this improved performance comes at much higher computational cost.

In conclusion, we have presented a proof of principle that a simple
time-frequency method could be used to detect GWs from bright EMRIs. A typical EMRI source could possibly be detected with SNR $> 6$ at a distance up to $\sim 2$ Gpc using this method. The method is computationally efficient in the sense that it takes only minutes to finish a search of EMRIs with one computer. Based on current estimates of the astrophysical rates \cite{jon04,freitag03}, tens of EMRIs could be detected each year by this technique. 

This method does not provide good parameter determination, but it could be used to detect the brightest sources as the first stage of a hierarchical search. The method provides some information about the frequency content and inspiral rate of an event which can be used to refine a subsequent matched filtering search. In practice, the EMRI detection problem will be made considerably more complicated by confusion with other sources in the LISA data, in particular confusion from white dwarf binaries. The time-frequency tracks of these other sources will look different to EMRIs. However, the tracks will overlap and a simple excess power method might not be able to distinguish multiple overlapping sources from one another. Further, in the current analysis, we have only considered a single `typical' EMRI signal, but the frequency and frequency evolution of other EMRIs will be different, which will change the detection statistics. Finally, the approximate quadrupole waveforms used in this analysis lack some of the multipole structure that we expect from true inspirals, which will also change our conclusions. More detailed discussion of these issues will be provided in a follow-up paper \cite{wen04}.

\ack
We thank Curt Cutler, Teviet Creighton, and Kip Thorne for critical discussions of this work, and Leor Barack for useful discussions and careful reading of the manuscript. LW thanks Max Planck Institut fuer Gravitationshphysik (Albert Einstein Institut) for support of this work. This work was supported in part by NASA grants NAG5-12384 and NAG5-10707 (JG).

\clearpage
\newpage

\end{document}